%% file: BPH-16-001_temp.tex
\pdfoutput=1

\documentclass[11pt,twoside,a4paper,cmspaper,final,collab]{cms-tdr}
\def\svnVersion{477173P}\def\cmsCernNoTag{CERN-EP-2018-131}\def\cmsCernDate{\today}\def\cmsMessage{Published in Physical Review Letters as \href{http://dx.doi.org/10.1103/PhysRevLett.121.141801}{\doi{10.1103/PhysRevLett.121.141801.}}}

\begin{document}\cmsNoteHeader{BPH-16-001}

\hyphenation{had-ron-i-za-tion}
\hyphenation{cal-or-i-me-ter}
\hyphenation{de-vices}
\RCS$HeadURL: svn+ssh://svn.cern.ch/reps/tdr2/papers/BPH-16-001/trunk/BPH-16-001.tex $
\RCS$Id: BPH-16-001.tex 477169 2018-10-05 15:37:49Z sleontsi $

\newlength\cmsFigWidth
\newlength\cmsTabSkip\setlength{\cmsTabSkip}{1ex}
\ifthenelse{\boolean{cms@external}}{\setlength\cmsFigWidth{0.98\columnwidth}}{\setlength\cmsFigWidth{0.65\textwidth}}
\ifthenelse{\boolean{cms@external}}{\providecommand{\cmsLeft}{top\xspace}}{\providecommand{\cmsLeft}{left\xspace}}
\ifthenelse{\boolean{cms@external}}{\providecommand{\cmsRight}{bottom\xspace}}{\providecommand{\cmsRight}{right\xspace}}
\newlength\cmsFigWidthFD
\ifthenelse{\boolean{cms@external}}{\setlength\cmsFigWidthFD{0.8\columnwidth}}{\setlength\cmsFigWidthFD{0.7\textwidth}}
\ifthenelse{\boolean{cms@external}}{\providecommand{\cmsLeft}{top\xspace}}{\providecommand{\cmsLeft}{left\xspace}}
\ifthenelse{\boolean{cms@external}}{\providecommand{\cmsRight}{bottom\xspace}}{\providecommand{\cmsRight}{right\xspace}}
\newcommand{\cPZJpsill}{\ensuremath{\cPZ\to\cPJgy\,\ell^+\ell^-}\xspace}
\newcommand{\cPZJpsimm}{\ensuremath{\cPZ\to\cPJgy\,\Pgmp\Pgmm}\xspace}
\newcommand{\cPZJpsiee}{\ensuremath{\cPZ\to\cPJgy\,\Pep\Pem}\xspace}
\newcommand{\cPZpsill}{\ensuremath{\cPZ\to\psi\,\ell^+\ell^-}\xspace}
\newcommand{\cPZpsimm}{\ensuremath{\cPZ\to\psi\,\Pgmp\Pgmm}\xspace}
\newcommand{\cPZpsiee}{\ensuremath{\cPZ\to\psi\,\Pep\Pem}\xspace}
\newcommand{\psill}{\ensuremath{\psi\,\ell^+\ell^-}\xspace}
\newcommand{\cPZfourmu}{\ensuremath{\cPZ\to\Pgmp\Pgmm\Pgmp\Pgmm}\xspace}
\newcommand{\psitomm}{\ensuremath{\psi\to\Pgmp\Pgmm}\xspace}
\newcommand{\Jpsitomm}{\ensuremath{\cPJgy\to\Pgmp\Pgmm}\xspace}
\newcommand{\lplm}{\ensuremath{\ell^+\ell^-}\xspace}
\newcommand{\mpmm}{\ensuremath{\Pgmp\Pgmm}\xspace}
\newcommand{\epem}{\ensuremath{\Pep\Pem}\xspace}

\cmsNoteHeader{BPH-16-001}

\title{Observation of the \texorpdfstring{$\cPZpsill$ decay in $\Pp\Pp$ collisions at $\sqrt{s}=13\TeV$}{Z to psi l+ l- decay in pp collisions at sqrt(s)= 13 TeV}}

\date{\today}

\abstract{
This Letter presents the observation of the rare $\cPZ$ boson decay $\cPZpsill$. Here, $\psi$ represents contributions from direct $\cPJgy$ and $\Pgy\to\cPJgy\,X$, $\lplm$ is a pair of electrons or muons, and the $\cPJgy$ meson is detected via its decay to $\mpmm$. The sample of proton-proton collision data, collected by the CMS experiment at the LHC at a center-of-mass energy of $13\TeV$, corresponds to an integrated luminosity of 35.9\fbinv. The signal is observed with a significance in excess of 5 standard deviations. After subtraction of the $\Pgy\to\cPJgy\,X$ contribution, the ratio of the branching fraction of the exclusive decay $\cPZJpsill$ to the decay $\cPZfourmu$ within a fiducial phase space is measured to be $\mathcal{B}(\cPZJpsill) / \mathcal{B}(\cPZfourmu) = 0.67\pm 0.18 \stat \pm 0.05 \syst.$
}

\hypersetup{
pdfauthor={CMS Collaboration},%
pdftitle={Observation of the Z -> ll psi decay in pp collisions at sqrt(s) = 13 TeV},%
pdfsubject={CMS},
pdfkeywords={CMS, physics, B-physics, rare decay}}

\maketitle

Although the $\cPZ$ boson was discovered more than 30 years ago~\cite{Arnison:1983mk}, only one exclusive decay channel with leptons, $\cPZ\to4\ell$~\cite{CMS:2012bw,Khachatryan:2016txa,Sirunyan:2017zjc,Aad:2014wra,Aad:2015rka}, has been observed apart from the dilepton final states. For radiative dilepton decays, $\cPZ\to\lplm\PGg$, where $\ell=\Pe$, \Pgm, experiments have reported yields consistent with the standard model, as well as upper limits on the branching fraction for anomalous production~\cite{Acton:1991dq}. No resonant structure in the four-lepton decay has yet been observed. The high rate of $\cPZ$ boson production at the CERN LHC facilitates the study of rare decay channels such as $\cPZ\to\text{V}\gamma$, $\cPZ\to\text{V}\lplm$, and $\cPZ\to\text{V}\,\text{V}$, where $\text{V}$ is a vector meson with $J^{PC} = 1^{--}$~\cite{Aad:2015sda,Aaboud:2017xnb}. In this paper, we present the observation of the decay of the $\cPZ$ boson to a final state with a $\cPJgy$ meson and two oppositely charged same-flavor leptons.

\begin{figure}[h!t]
\centering
{\includegraphics[width=\cmsFigWidthFD]{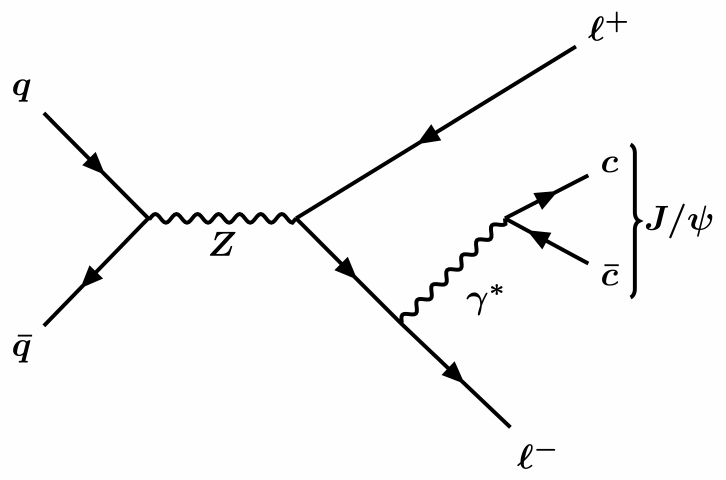}}
\caption{Leading-order Feynman diagram for the production of the \cPZ\ boson and its decay in the $\cPZJpsill$ channel.}
\label{fig:feynman}
\end{figure}

The $\cPZ\to\text{V}\,\lplm$ process has been described and studied in various theoretical papers~\cite{Bergstrom:1990uu,Fleming:1993fq,Fleming:1994iu,Li:2010xu,Braaten:1993mp,Ernstrom:1996aa,Bergstrom:1990bu}. For the case where $\text{V}=\cPJgy$, the branching fraction $\mathcal{B}(\cPZJpsill)$ is calculable within the standard model. The dominant diagram is the quantum electrodynamics radiative process illustrated in Fig.~\ref{fig:feynman}, with the $\gamma^*$--$\text{V}$ transition strength derived from the measured $\text{V}\to\lplm$ electromagnetic decays~\cite{PDG2016}. The theoretical estimates of the branching fraction cover the range $(6.7$--$7.7)\times 10^{-7}$~\cite{Bergstrom:1990uu,Fleming:1993fq}. Although this branching fraction is small, the dileptons and vector meson in the final state offer a clean signature. The measurement of this branching fraction is valuable for the calculation of the fragmentation function for a virtual photon to split into a $\cPJgy$ meson. Rare Higgs boson decays, such as those to quarkonia~\cite{Doroshenko:1987nj,Gao:2014xlv}, will become accessible in the future, making it possible to search for nonstandard model signatures in these decays, including, e.g., anomalous couplings or new exotic light states~\cite{Gonzalez-Alonso:2014rla}. Accurate knowledge of potential backgrounds from $\cPZ$ decays to quarkonia will be essential for these measurements.

This analysis uses proton-proton $(\Pp\Pp)$ collision data recorded by the CMS experiment at a center-of-mass energy of $13\TeV$, corresponding to an integrated luminosity of $35.9\fbinv$. We report the observation of the \cPZpsill decay channel, where $\psi$ represents the contributions from direct $\cPJgy$ and $\cPJgy$ mesons from $\Pgy$ decays, and the $\cPJgy$ is detected via its $\mpmm$ decay channel. We measure the ratio of the branching fraction of this decay to that of the \cPZfourmu decay, to take advantage of a partial cancellation of systematic uncertainties.

The central feature of the CMS apparatus is a superconducting solenoid of 6\unit{m} internal diameter, providing a magnetic field of 3.8\unit{T}. Within the solenoid volume are a silicon pixel and strip tracker, a lead tungstate crystal electromagnetic calorimeter (ECAL), and a brass and scintillator hadron calorimeter, each composed of a barrel and two endcap sections. Muons are detected in gas-ionization chambers embedded in the steel flux-return yoke outside the solenoid, in the pseudorapidity range $\abs{\eta}<2.4$~\cite{Chatrchyan:2012xi}. Electrons are reconstructed using information from the ECAL and the tracker, in the $\abs{\eta}<2.5$ range~\cite{Khachatryan:2015hwa}. A more detailed description of the CMS detector, together with a definition of the coordinate system used and the relevant kinematic variables, can be found in Ref.~\cite{Chatrchyan:2008zzk}.

Events of the \cPZfourmu process are simulated with the next-to-leading-order Monte Carlo (MC) generator \POWHEG~\cite{Frixione:2007vw}, interfaced with \PYTHIA 8.175~\cite{Sjostrand:2007gs,Sjostrand:2014zea} with parameters set by the CUETP8M1 tune~\cite{Khachatryan:2015pea} for parton showering, hadronization, and the underlying event.  The parton distribution functions are taken from the NNPDF 3.0~\cite{Ball:2014uwa} set. For the \cPZJpsill signal we use \PYTHIA 8.175 (same tune as for \cPZfourmu) to simulate the production of $\cPZ$ bosons, with an unpolarized phase-space model for the \cPZJpsill decay.  Matrix-element effects are evaluated by comparison with data and treated as systematic uncertainties. The detector response is simulated with a model of the CMS detector implemented in the \GEANTfour package~\cite{Agostinelli:2002hh}. We measure the fiducial branching fraction restricted to a region of phase space covered by the acceptance of the measurement, as described below.

The trigger and offline selection criteria closely follow the previous CMS analysis of $\cPZ\to4\ell$ decays~\cite{CMS:2012bw,Khachatryan:2016txa,Sirunyan:2017zjc}. Triggers requiring one, two, or three charged leptons, with varying \pt requirements, are used.  The combined efficiency of the triggers, within the acceptance of this analysis defined below, is greater than 99\%.

Among the multiple $\Pp\Pp$ collisions within the time resolution of the data acquisition, the primary vertex is taken to be the reconstructed vertex with the largest sum of $\pt^2$ over the physics objects in the event. These objects include jets, clustered using the anti-\kt jet finding algorithm~\cite{Cacciari:2008gp,Cacciari:2011ma} with the tracks assigned to the vertex as inputs, and the associated missing transverse momentum, taken as the negative vector sum of the \ptvec of those jets.  The primary vertex is required to lie within 24\unit{cm} of the center of the detector along the beam axis and 2\unit{cm} perpendicular to that axis. Charged particle tracks associated with vertices other than the primary vertex are ignored.

We require all lepton candidate trajectories to pass within 1\,(0.5)\unit{cm} of the primary vertex in the direction along (perpendicular to) the beam axis. The lepton candidates from $\cPZ$ boson decay are required to be isolated from the hadronic activity in the event. To satisfy this requirement, the scalar sum of transverse energy deposits in the calorimeters and the $\pt$ of tracks is computed in a cone of radius $\Delta R\equiv\sqrt{\smash[b]{(\Delta\eta)^2+(\Delta\phi)^2}}=0.3$ in $\eta$--$\phi$ around the lepton trajectory, where $\phi$ is the azimuthal angle in radians. The sum is corrected for other leptons from $\cPZ$ boson decay that fall within the isolation cone and for the average hadronic activity in an event. The ratio of this corrected sum to the lepton \pt is required to be smaller than $0.35$. Leptons are required to be separated by $\Delta R>0.02\ (0.05)$ for same- (different-) flavor pairs.

We select events with two oppositely charged reconstructed muons consistent with the dimuon decay of a $\cPJgy$ meson that, in combination with two additional oppositely charged electrons or muons (which we refer to as prompt leptons, $\ell$), are consistent with the decay \cPZpsill. Specifically, the invariant mass of the $\psi$ muon pair must satisfy $2.6<m_{\mpmm}<3.6\GeV$ and that of the four leptons must satisfy $\abs{m_{\mpmm\lplm}-m_{\cPZ}}<25\GeV$, where $m_{\cPZ}=91.2\GeV$~\cite{PDG2016}. Each of the muons from $\cPJgy$ decay are required to have $\pt>3.5\GeV$ and $\abs{\eta}<2.4$, and the $\pt$ of the $\cPJgy$ candidate must exceed $8.5\GeV$. We require the highest- and second-highest-$\pt$ prompt leptons to have $\pt>30$ and $15\GeV$, respectively, satisfy $\abs{\eta}<2.5\,(2.4)$ for $\ell=\Pe\,(\PGm)$, and have a dilepton invariant mass $m_{\lplm}<80\GeV$.  The lepton $\pt$ thresholds ensure high trigger efficiency, and the invariant mass requirement suppresses the background from events in which a dilepton from $\cPZ$ boson decay is combined with a dimuon from an uncorrelated $\cPJgy$ decay or a nonresonant muon pair.

The four leptons, and separately the two muons from the $\cPJgy$ decay, are fitted to common vertices, with each vertex fit required to have a $\chi^2$ probability greater than $5\%$. The significance of the three-dimensional impact parameter relative to the primary vertex is required to satisfy $\abs{d_\text{IP}/\sigma_\text{IP}}<4$ for each lepton, where $d_{\text{IP}}$ is the signed distance of closest approach of the lepton track to the primary vertex and $\sigma_\text{IP}$ is the associated uncertainty.

\begin{figure}[h!t]
\centering
{\includegraphics[width=\cmsFigWidth]{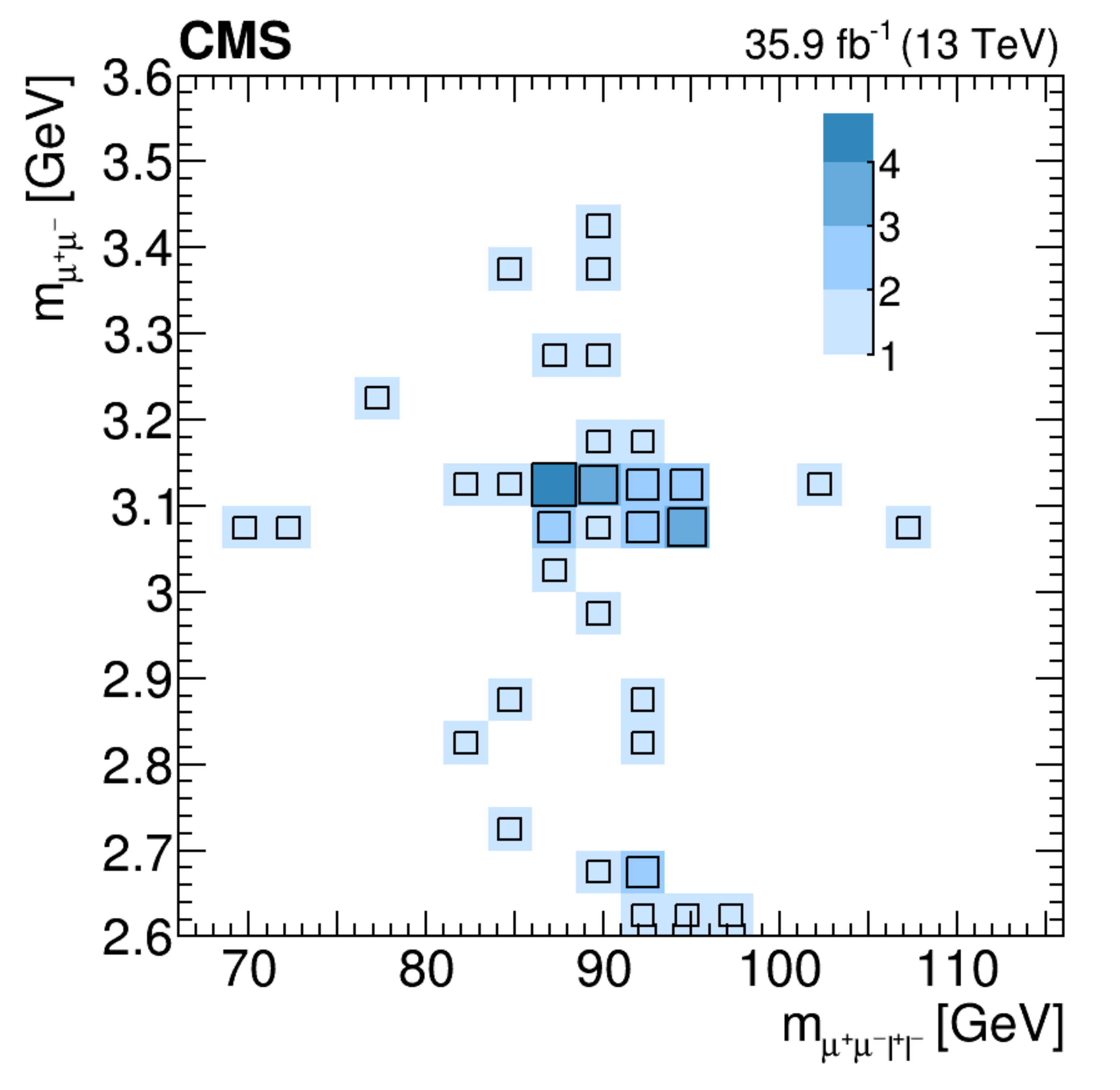}}
\caption{Distribution of invariant masses $m_{\mpmm}$ vs. $m_{\mpmm\lplm}$ for the selected candidates. The values in the legend give the numbers of candidates per bin, which are also indicated by the sizes of the open black boxes.}
\label{fig:AllEventsCombined}
\end{figure}

Following the application of the selection criteria described above, $29\,(18)$ events remain in the $\psi\,\mpmm\,(\psi\,\epem)$ sample. Figure~\ref{fig:AllEventsCombined} shows a two-dimensional plot of the $\mpmm$ versus $\mpmm\lplm$ invariant masses for the candidate events. The signal appears as a concentration of events in the overlap region of the $\cPJgy$ meson and $\cPZ$ boson masses. The events outside the central cluster along the $\cPZ$ boson mass band indicate contributions from the $\cPZ\to\text{(continuum~\Pgmp\!\Pgmm)}\,\lplm$ decay, and along the $\cPJgy$ meson mass band, nonresonant $\cPJgy\,\lplm$ production.

We measure the branching fraction of the \cPZpsill decay mode relative to that of \cPZfourmu. The selection criteria for the \cPZfourmu events follow Ref.\ \cite{Sirunyan:2017zjc}; here the required mass ranges of the two oppositely charged muon pairs are $4\,(40)<m(\mpmm)<80\GeV$, where the $40\GeV$ threshold applies to the pair with the larger invariant mass.

\begin{figure}[h!t]
\centering
{\includegraphics[width=0.49\columnwidth]{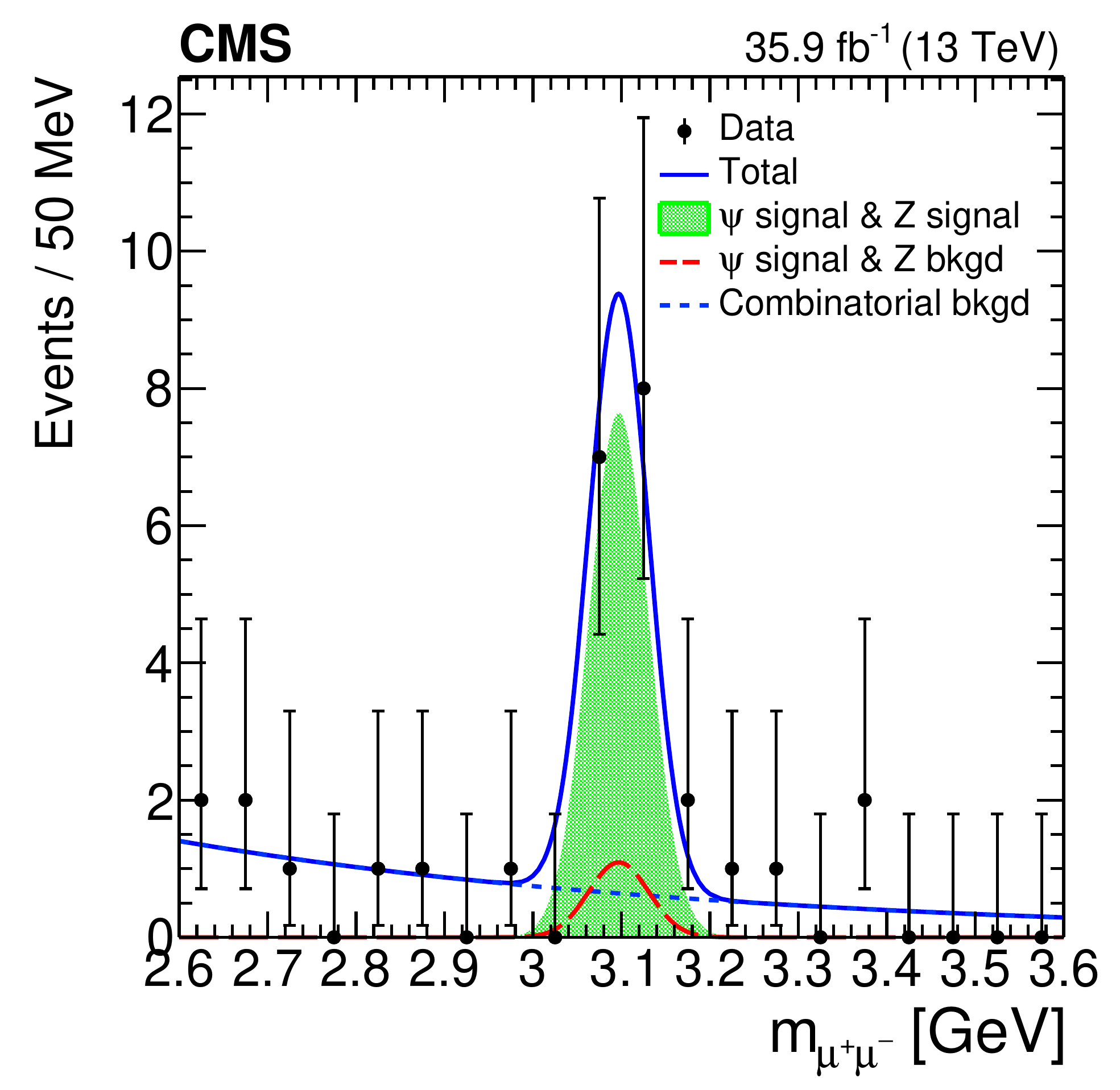}}
{\includegraphics[width=0.49\columnwidth]{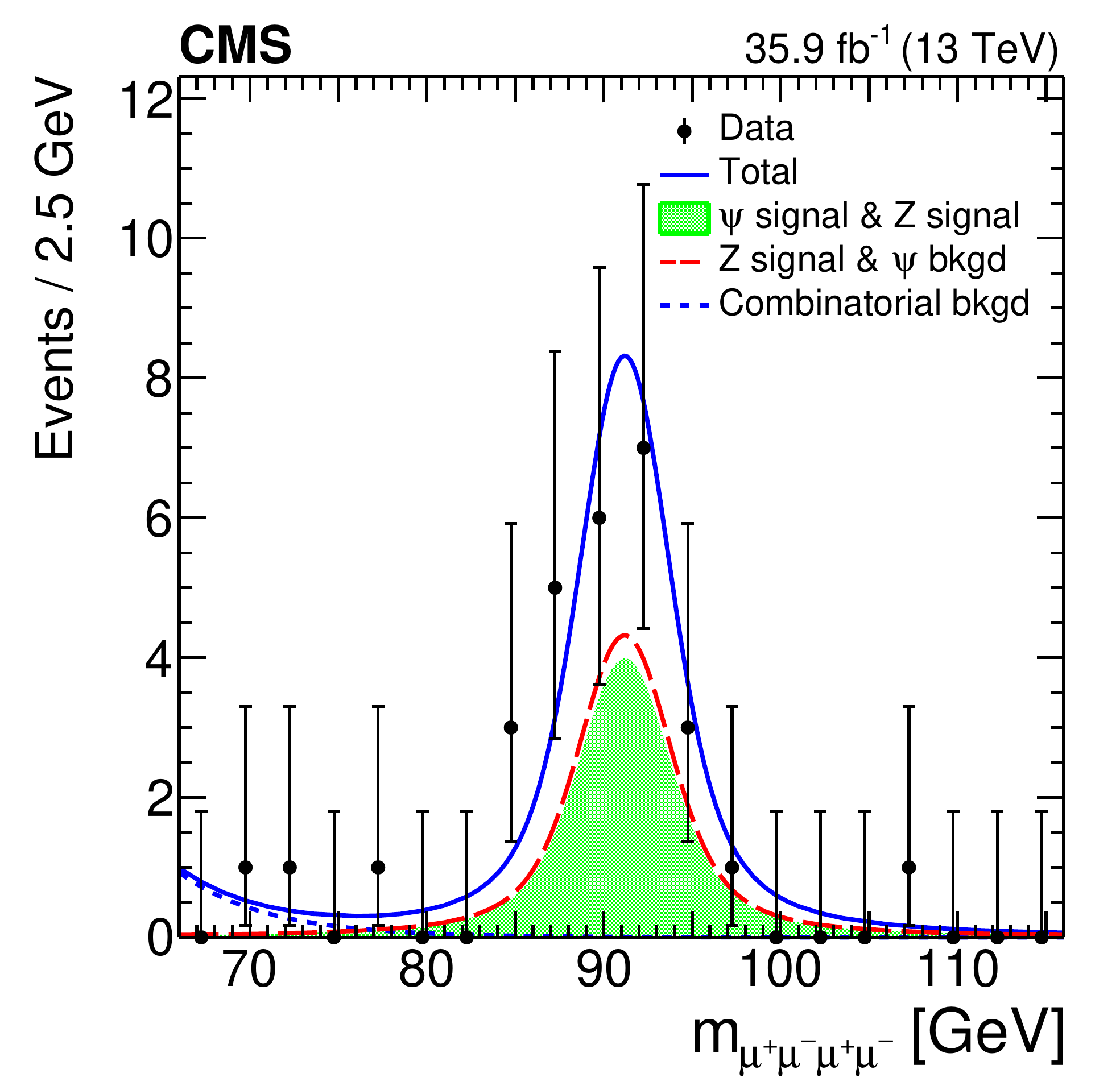}}\\
{\includegraphics[width=0.49\columnwidth]{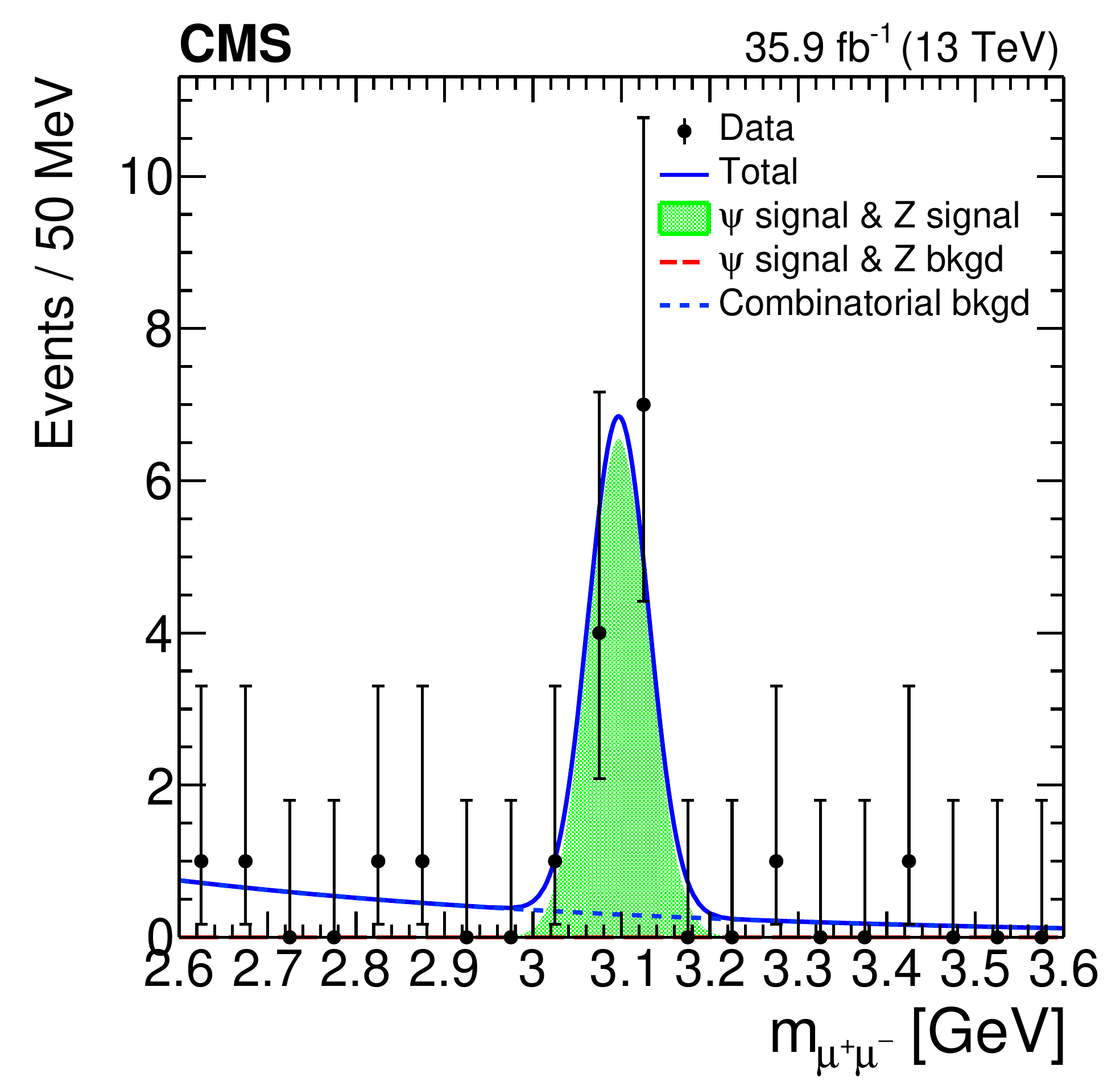}}
{\includegraphics[width=0.49\columnwidth]{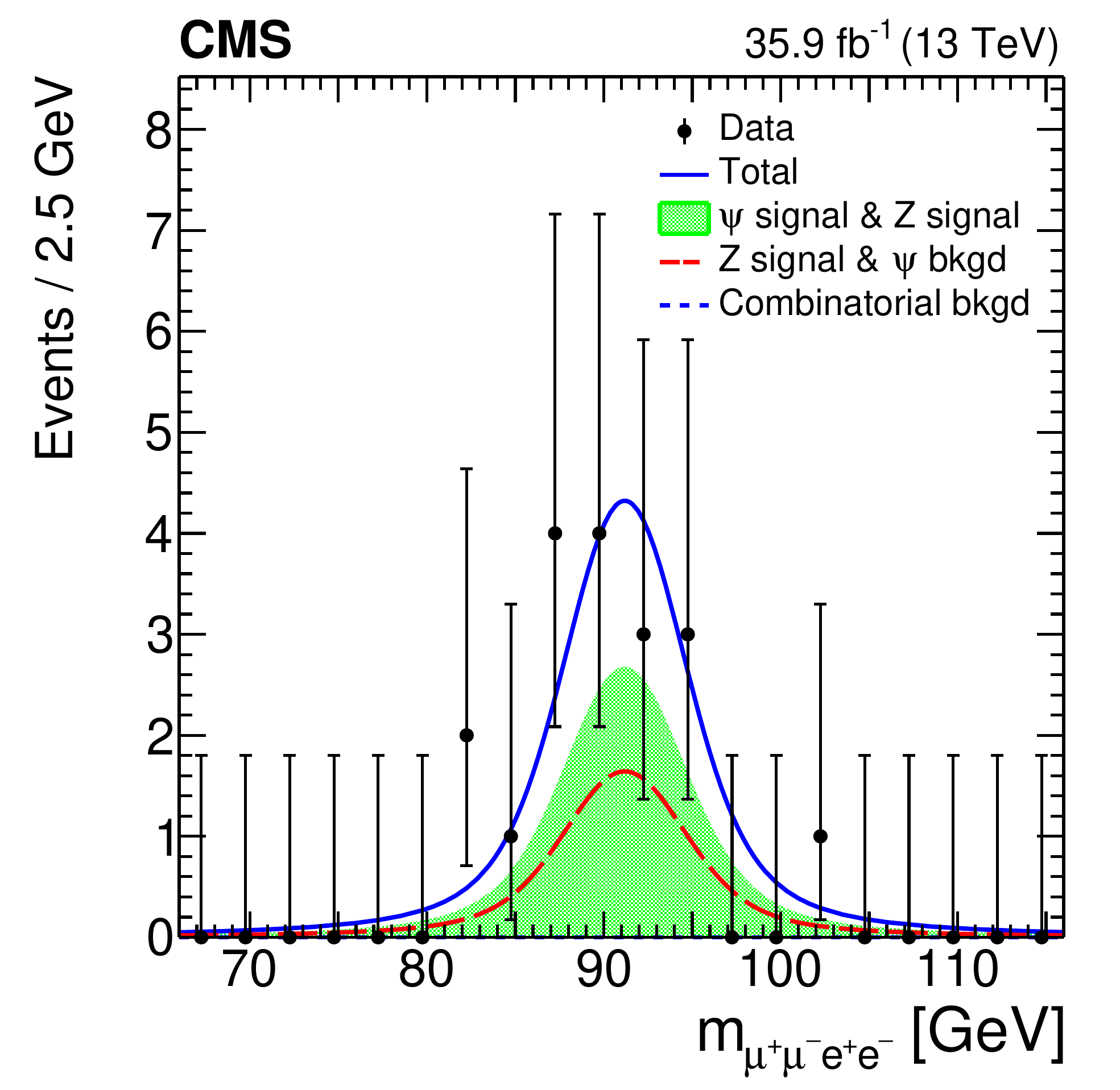}}
\caption{Invariant mass distributions for the $\psi$ muon pairs (left) and for \psill (right), for \cPZpsimm (upper) and \cPZpsiee (lower) candidates. In each histogram the
data are represented by the points, with the vertical bars showing the statistical uncertainties, and the solid curve is the overall fit to the data. The shaded region corresponds to the signal yield, while the long-dashed lines are the $\psi$ meson signal from the $\cPZ$ boson background (left) and the $\cPZ$ boson signal from the $\psi$ meson background (right). The short-dashed line represents the combinatorial background.
}
\label{fig:AllEventsZpsi}
\end{figure}

The signal yield is obtained from unbinned extended maximum-likelihood fits~\cite{Barlow:1990vc} of the distributions in the two invariant mass variables $m_{\mpmm}$ and $m_{\mpmm\lplm}$, separately for the dimuon and dielectron channels.  The probability density function (pdf) is a sum of four terms, each of which is a yield parameter multiplying a component pdf of the form $f(m_{\mpmm})g(m_{\mpmm\lplm})$.  The four terms account for the $\cPZ\to\psi\,\lplm$ signal and the backgrounds from: $\cPZ\to\lplm$ accompanied by nonresonant $\mpmm$; nonresonant $\cPJgy\lplm$; and nonresonant $\mpmm\lplm$. The pdf for the $\cPJgy\to\mpmm$ invariant mass distribution is a Gaussian function of $m_{\mpmm}$ with the mean fixed to the $\cPJgy$ meson mass~\cite{PDG2016} and the width as a free parameter of the fit. The $\cPZ\to\mpmm\lplm$ pdf is a Breit--Wigner function of $m_{\mpmm\lplm}$ with its central value and width fixed to the mass and width of the $\cPZ$ boson~\cite{PDG2016}, convolved with a Gaussian function whose width is a free parameter.  The pdfs for the continuum background in each dimension of the fit, representing backgrounds that are both peaking and nonpeaking in the orthogonal dimension, are exponential functions with free decay parameters.  The projections in each variable are shown in Fig.~\ref{fig:AllEventsZpsi}, along with the pdf components resulting from the fits.

The yields resulting from the fit are $13.0\pm 3.9$ events for the $\cPZpsimm$ mode and $11.2\pm 3.4$ events for $\cPZpsiee$, where the uncertainties are statistical only. The yields of the two decay modes agree within uncertainties, as expected, since the reconstruction efficiencies of the prompt electrons and muons in this \pt range are similar. The \cPZfourmu reference signal is extracted with a separate extended unbinned maximum-likelihood fit to the $m_{\mpmm\mpmm}$ distribution, using the same parametrization as for \cPZpsill. The fit yields $250\pm 20$ events.

We evaluate the signal significance for both $\psi\,\mpmm$ and $\psi\,\epem$ by generating random pseudo-experiments with dimuon and four-lepton invariant mass distributions drawn from the background-only pdf and then fitted with the background-only and signal-plus-background hypotheses. From the pseudo-experiments the likelihood ratio of the two hypotheses is calculated and compared with the likelihood ratio of the data. Taking into account the systematic uncertainties (discussed below), the background-only hypothesis is excluded at $4.0$ and $4.3$ standard deviations for $\psi\,\mpmm$ and $\psi\,\epem$, respectively. The combination of the two significances based on the Fisher formalism~\cite{ref:FisherComb} results in the observation of the \cPZpsill decay mode with a significance of $5.7$ standard deviations.

From the observed signal yield we compute a ratio of branching fractions defined over the fiducial phase space of the measurement defined in Table~\ref{tab:phasespaceMCtruth}.  The entries consist of the kinematical requirements of the event selection given above, plus the additional requirement $m_{\ell^{+}\ell^{-}}>40\GeV$ for the \cPZpsill candidates, which is added to match the selection of the \cPZfourmu candidates and to avoid regions of the decay phase space in which the acceptance is steeply falling.  This requirement removes 2\,(0) events from the \cPZpsiee (\cPZpsimm) sample, and 0.95 events from the fitted \cPZpsiee yield. The ratio of the fiducial branching fractions for lepton flavor $\ell$ is
\ifthenelse{\boolean{cms@external}}{
\begin{widetext}
\begin{equation}
\mathcal{R}_{\cPJgy\lplm} = \frac{\mathcal{B}(\cPZJpsill)}{\mathcal{B}(\cPZfourmu)} = \left(\frac{1}{2}\sum_{{l_i}=\Pgm,\Pe}{\frac{N_{\cPZ\to\cPJgy\ell_i^+\ell_i^-}}{\epsilon_{\cPZ\to\cPJgy\ell_i^+\ell_i^-}}}\right)
\frac{\epsilon_{\cPZfourmu}}{N_{\cPZfourmu}}\frac{1}{\mathcal{B}(\Jpsitomm)},
\label{eq:brformula}
\end{equation}
\end{widetext}
}{
\begin{equation}
\mathcal{R}_{\cPJgy\lplm} = \frac{\mathcal{B}(\cPZJpsill)}{\mathcal{B}(\cPZfourmu)} = \left(\frac{1}{2}\sum_{{l_i}=\Pgm,\Pe}{\frac{N_{\cPZ\to\cPJgy\ell_i^+\ell_i^-}}{\epsilon_{\cPZ\to\cPJgy\ell_i^+\ell_i^-}}}\right)\frac{\epsilon_{\cPZfourmu}}{N_{\cPZfourmu}}\frac{1}{\mathcal{B}(\Jpsitomm)},
\label{eq:brformula}
\end{equation}
}
where the branching fraction $\mathcal{B}(\Jpsitomm)=(5.961\pm 0.033)\%$~\cite{PDG2016}, $N_{\cPZ\to\cPJgy\ell_i^+\ell_i^-}$ is the signal yield excluding the $\Pgy\to\cPJgy\,X$ contribution, and $N_{\cPZfourmu}$ is the reference-channel yield. The experimental efficiencies to reconstruct events within the fiducial phase space are determined from simulation; combined with the trigger efficiencies given above they are $\epsilon_{\cPZJpsimm}=81\%$, $\epsilon_{\cPZJpsiee}=80\%$, and $\epsilon_{\cPZfourmu}=81\%$.

\begin{table}[h!t]
\centering
\topcaption{\label{tab:phasespaceMCtruth} Definition of the fiducial phase space for the measurement of the ratio of branching fractions. Here, $\ell$ refers to a prompt muon or electron from the signal decay, or to either of the two muons from the higher invariant-mass pair in the reference-channel decay, and \Pgm\ refers to a $\cPJgy$ daughter or a member of the lower invariant-mass pair in the reference-channel decay.  The symbol $\ell_1$ ($\ell_2$) refers to the prompt lepton having the higher (lower) value of \pt. The $\pt^{\cPJgy}$ threshold is applied to the signal and the $m({\PGm^\pm\Pgm^\mp})$ requirement to the reference channel.
}
\begin{scotch}{l}
Fiducial requirement\\
\hline\\[-2.0ex]
$40<m_{\ell^{+}\ell^{-}}<80\GeV$\\
$\abs{\eta(\text{electrons})}<2.5$,\quad  $\abs{\eta(\text{muons})}<2.4$ \\
$\pt(\ell_1,\ell_2,\PGm,\PGm)>(30,15,3.5,3.5)\GeV$\\
Signal:  $\pt^{\cPJgy}>8.5\GeV$\\
Reference channel:  $4<m({\PGm^\pm\Pgm^\mp})<80\GeV$\\
\end{scotch}
\end{table}

Calculated contributions from $\Pgy\to\cPJgy\,X$ decays, the dominant feed-down source of $\cPJgy$ mesons,  are subtracted from the signal yields, since the natural width of the $\cPZ$ boson does not allow the separation of the process $\Pgy\to\cPJgy\,X$ from direct $\cPJgy$ production. The predicted production ratio of $\cPZJpsill$ to $\cPZ\to\Pgy\,\ell^{+}\ell^{-}$ is $3.5$~\cite{Fleming:1993fq}. Taking into account the branching fraction of \Pgy\ to $\cPJgy\,X$~\cite{PDG2016}, the ratio of $N\left(\cPZJpsill\right)$ to $N\left(\cPZ \to \Pgy[\to\cPJgy\,X]\lplm \right)$ is $5.7\pm 0.1$. Using this scale factor, we subtract $1.9$ $(1.7)$ events from the $N_{\cPZpsimm} (N_{\cPZpsiee})$ yield, considering them as $\cPJgy$ events from \Pgy\ meson decays.

Since the signal and reference-channel events are recorded with the same triggers, and the topologies of the selected events are similar, many systematic uncertainties cancel in the ratio. The uncertainties in $\mathcal{R}_{\cPJgy\lplm}$ are shown for the two signal decay modes in columns $2$ and $3$ of Table~\ref{tab:systematics}, and are combined in quadrature as uncorrelated, unless stated otherwise, in column $4$.

Systematic uncertainties arising from the choice of fit model are calculated by varying the pdfs used for the signal ($\cPZ$ and $\cPJgy$) and combinatorial background. Substitution of a double-Gaussian function for the $\cPZ$ boson signal leads to differences in the signal yields of $0.02$, $0.05$, and $1.88$ events in $\cPZpsimm$, $\cPZpsiee$, and $\cPZfourmu$, respectively. The corresponding changes from using a first-order polynomial instead of an exponential function for the $\cPZ$ boson combinatorial background are $0.9$, $0.1$, and $0.4$ events.

A similar approach was followed for the $\cPJgy$ meson signal and background pdfs. The maximum difference observed in the signal yields resulting from the substitution of the sum of a double-Gaussian and a Crystal Ball~\cite{Oreglia:1980cs} function for the signal pdf is $0.6$ events for the $\psi\,\mpmm$ and $0.2$ events for the $\psi\,\epem$ final state. The background pdf was replaced by a first-order polynomial to estimate the background model uncertainty, where a difference of $0.2$ events is found in both decay modes.

To measure the uncertainty from the fitting procedure, $1000$ random pseudo-samples were generated with the number of events of each drawn from a Poisson distribution having a mean equal to the number of events observed in the data.  The absolute value of the average deviation of the fit yields from the nominal yield is taken as the systematic uncertainty.

The reconstruction efficiencies of the muons from $\cPJgy$ decay and prompt leptons (electrons and muons) are checked with $\cPZ\to\mpmm$, $\cPZ\to\epem$, and $\cPJgy\to\mpmm$ decay data using the ``tag-and-probe" method~\cite{Khachatryan:2010xn,Chatrchyan:2012xi}, as functions of the lepton $\eta$ and $\pt$. To calculate the systematic uncertainty in $\mathcal{R}_{\cPJgy\lplm}$, these efficiencies are varied within their uncertainty, with the uncertainties from the lepton efficiencies treated as correlated in the ratio. In addition, we assign an uncertainty associated with the finite number of MC signal and reference-channel events used to obtain the reconstruction efficiencies.

We test the three-body $\cPZ$ boson decay model implemented in the MC simulation by comparing distributions from the simulation with those from signal-weighted data, obtained from the fit model by the $_{s}\mathcal{P}lot$ method~\cite{Pivk:2004ty}.  The most sensitive observables were found to be the azimuthal separation between the $\cPJgy$ candidate and the highest- and second-highest-\pt prompt leptons.  We apply the observed shape differences to the simulation and reevaluate the reconstruction efficiency to extract the decay model uncertainty.

The uncertainty in the fraction of $\cPJgy$ events that potentially originate from $\Pgy$ is propagated from the uncertainty of the $N\left(\cPZJpsill\right)$ to $N\left(\cPZ \to \Pgy[\to\cPJgy\,X]\lplm \right)$ ratio.

The total systematic uncertainty of 7.6\% for $\mathcal{R}_{\cPJgy\lplm}$ is calculated by adding the sources of uncertainty given in the last column of Table~\ref{tab:systematics} in quadrature.

\begin{table}[h!t]
  \centering
  \topcaption{The contributions to the systematic uncertainty in the ratio of branching fractions for the prompt muon, prompt electron, and combined samples, in percent.  The last row gives the sum in quadrature of all components.}
  \begin{scotch}{lccc}
 Source of uncertainty             & $\mathcal{R}_{\cPJgy\mpmm}$          & $\mathcal{R}_{\cPJgy\epem}$        & $\mathcal{R}_{\cPJgy\lplm}$     \\ \hline
  $\cPZ$ boson signal shape        & $0.8$                                & $0.8$                              & $0.8$                           \\
  $\cPZ$ boson background shape    & $6.9$                                & $0.5$                              & $3.7$                           \\
  $\cPJgy$ meson signal shape      & $4.8$                                & $2.0$                              & $2.8$                           \\
  $\cPJgy$ meson background shape  & $1.5$                                & $1.5$                              & $1.1$                           \\
  Fit procedure                    & $3.0$                                & $8.4$                              & $4.2$                           \\
  Reconstruction efficiency        & $0.9$                                & $5.9$                              & $4.0$                           \\
  MC sample size                   & $0.7$                                & $0.8$                              & $0.5$                           \\
  $\cPZ$ boson decay model         & $0.7$                                & $1.6$                              & $0.8$                           \\
  \Pgy\ feed-down                  & $0.3$                                & $0.3$                              & $0.3$                           \\ [\cmsTabSkip]
  Total                            & $9.2$                                & $10.8$                             & $7.6$                           \\
  \end{scotch}
\label{tab:systematics}
\end{table}

After subtracting the \Pgy\ feed-down we extract from Eq.~(\ref{eq:brformula}) the branching fraction ratio $\mathcal{R}_{\cPJgy\,\lplm}$, for the phase-space region defined in Table~\ref{tab:phasespaceMCtruth}:
\begin{equation*}
\mathcal{R}_{\cPJgy\,\lplm} = 0.67\pm 0.18\stat \pm 0.05\syst.
\end{equation*}

Assuming that the factors applied to extrapolate the signal and reference-channel branching fractions from the phase space defined in Table~\ref{tab:phasespaceMCtruth} to the full phase space approximately cancel in the ratio, we use the measured value of $\mathcal{B}(\cPZfourmu)=(1.20\pm 0.08)\times 10^{-6}$~\cite{Sirunyan:2017zjc} for $m(\mpmm)>4\GeV$ to obtain an estimate for $\mathcal{B}(\cPZJpsill)$ of $8\times 10^{-7}$. This estimate is consistent with standard model predictions of $(6.7\pm 0.7)\times 10^{-7}$~\cite{Bergstrom:1990uu} and $7.7\times 10^{-7}$~\cite{Fleming:1993fq}.

The factors that extrapolate the fiducial measurements to the full phase space depend on the $\cPZ$ boson decay matrix element, which determines the angular distributions of the muons coming from the $\psi$ meson and the prompt leptons. Computing those factors assuming that the $\psi$ is transversely or longitudinally polarized in the helicity frame ($\lambda_\theta = \pm 1$)~\cite{Faccioli:2010kd} leads to a full phase space branching fraction ratio that differs by less than $25\%$ from the unpolarized result.

In summary, a new decay mode of the $\cPZ$ boson into a $\psi$ meson, where $\psi$ represents the contributions from direct $\cPJgy$ and $\Pgy\to\cPJgy\,X$, and an additional pair of leptons (muons or electrons), is observed with a statistical significance greater than 5 standard deviations. Using data from proton-proton collisions collected with the CMS detector at $\sqrt{s}=13\TeV$, corresponding to an integrated luminosity of 35.9\fbinv, $13.0\pm 3.9$ events of the $\cPZpsimm$ and $11.2\pm 3.4$ events of the $\cPZpsiee$ decay are obtained. This is the first observed $\cPZ$ boson decay to a vector meson and two oppositely charged same-flavor leptons. The ratio of the branching fraction for this decay to the one for the reference channel $\cPZfourmu$ in the fiducial phase space of the measurement, as defined in Table~\ref{tab:phasespaceMCtruth}, after subtracting the \Pgy\ feed-down, is $\mathcal{R}_{\cPJgy\,\lplm} = 0.67\pm 0.18\stat \pm 0.05\syst$. Using the known branching fraction for $\cPZfourmu$ results in a branching fraction for $\cPZJpsill$ consistent with standard model predictions.

\begin{acknowledgments}
We congratulate our colleagues in the CERN accelerator departments for the excellent performance of the LHC and thank the technical and administrative staffs at CERN and at other CMS institutes for their contributions to the success of the CMS effort. In addition, we gratefully acknowledge the computing centers and personnel of the Worldwide LHC Computing Grid for delivering so effectively the computing infrastructure essential to our analyses. Finally, we acknowledge the enduring support for the construction and operation of the LHC and the CMS detector provided by the following funding agencies: BMWFW and FWF (Austria); FNRS and FWO (Belgium); CNPq, CAPES, FAPERJ, FAPERGS, and FAPESP (Brazil); MES (Bulgaria); CERN; CAS, MoST, and NSFC (China); COLCIENCIAS (Colombia); MSES and CSF (Croatia); RPF (Cyprus); SENESCYT (Ecuador); MoER, ERC IUT, and ERDF (Estonia); Academy of Finland, MEC, and HIP (Finland); CEA and CNRS/IN2P3 (France); BMBF, DFG, and HGF (Germany); GSRT (Greece); NKFIA (Hungary); DAE and DST (India); IPM (Iran); SFI (Ireland); INFN (Italy); MSIP and NRF (Republic of Korea); LAS (Lithuania); MOE and UM (Malaysia); BUAP, CINVESTAV, CONACYT, LNS, SEP, and UASLP-FAI (Mexico); MBIE (New Zealand); PAEC (Pakistan); MSHE and NSC (Poland); FCT (Portugal); JINR (Dubna); MON, RosAtom, RAS, RFBR, and NRC KI (Russia); MESTD (Serbia); SEIDI, CPAN, PCTI, and FEDER (Spain); Swiss Funding Agencies (Switzerland); MST (Taipei); ThEPCenter, IPST, STAR, and NSTDA (Thailand); TUBITAK and TAEK (Turkey); NASU and SFFR (Ukraine); STFC (United Kingdom); DOE and NSF (USA). \end{acknowledgments}

\bibliography{auto_generated}
\cleardoublepage \appendix\section{The CMS Collaboration \label{app:collab}}\begin{sloppypar}\hyphenpenalty=5000\widowpenalty=500\clubpenalty=5000\input{BPH-16-001-authorlist.tex}\end{sloppypar}
\end{document}

%% file: BPH-16-001-authorlist.tex
\vskip\cmsinstskip
\textbf{Yerevan Physics Institute, Yerevan, Armenia}\\*[0pt]
A.M.~Sirunyan, A.~Tumasyan
\vskip\cmsinstskip
\textbf{Institut f\"{u}r Hochenergiephysik, Wien, Austria}\\*[0pt]
W.~Adam, F.~Ambrogi, E.~Asilar, T.~Bergauer, J.~Brandstetter, E.~Brondolin, M.~Dragicevic, J.~Er\"{o}, A.~Escalante~Del~Valle, M.~Flechl, R.~Fr\"{u}hwirth\cmsAuthorMark{1}, V.M.~Ghete, J.~Hrubec, M.~Jeitler\cmsAuthorMark{1}, N.~Krammer, I.~Kr\"{a}tschmer, D.~Liko, T.~Madlener, I.~Mikulec, N.~Rad, H.~Rohringer, J.~Schieck\cmsAuthorMark{1}, R.~Sch\"{o}fbeck, M.~Spanring, D.~Spitzbart, A.~Taurok, W.~Waltenberger, J.~Wittmann, C.-E.~Wulz\cmsAuthorMark{1}, M.~Zarucki
% ...
\vskip\cmsinstskip
% ...
1:  Also at Vienna University of Technology, Vienna, Austria\\
% ...


%% file: BPH-16-001_temp.bbl
\providecommand{\href}[2]{#2}\begingroup\raggedright\begin{thebibliography}{10}%
\makeatletter
\providecommand{\hrefCMSnoop }[0]{\@secondoftwo}%
\makeatother
\providecommand{\doi}{\texttt{doi:}\begingroup \urlstyle{tt}\Url}

\bibitem{Arnison:1983mk}
\hrefCMSnoop {}{{UA1} Collaboration, ``{Experimental observation of lepton
  pairs of invariant mass around $95\,\mathrm{GeV}/\mathrm{c}^2$ at the CERN
  SPS collider}'',} \textit{ Phys. Lett. B} \textbf{ 126} (1983) 398,
\href{http://dx.doi.org/10.1016/0370-2693(83)90188-0}{\doi{10.1016/0370-2693(83)90188-0}}.

\bibitem{CMS:2012bw}
\hrefCMSnoop {}{{CMS Collaboration}, ``{Observation of \Z decays to four
  leptons with the CMS detector at the LHC}'',} \textit{ JHEP} \textbf{ 12}
  (2012) 034,
  \href{http://dx.doi.org/10.1007/JHEP12(2012)034}{\doi{10.1007/JHEP12(2012)034}},
\href{http://www.arXiv.org/abs/1210.3844}{\texttt{arXiv:1210.3844}}.

\bibitem{Khachatryan:2016txa}
\hrefCMSnoop {}{{CMS Collaboration}, ``{Measurement of the ZZ production cross
  section and Z $\to \ell^+\ell^-\ell'^+\ell'^-$ branching fraction in pp
  collisions at $\sqrt{s}=13\,\mathrm{TeV}$}'',} \textit{ Phys. Lett. B}
  \textbf{ 763} (2016) 280,
  \href{http://dx.doi.org/10.1016/j.physletb.2016.10.054}{\doi{10.1016/j.physletb.2016.10.054}},
  \href{http://www.arXiv.org/abs/1607.08834}{\texttt{arXiv:1607.08834}}.
[Erratum: \DOI{10.1016/j.physletb.2017.09.030}].

\bibitem{Sirunyan:2017zjc}
\hrefCMSnoop {}{{CMS Collaboration}, ``{Measurements of the $\mathrm {p}\mathrm
  {p}\rightarrow \mathrm{Z}\mathrm{Z}$ production cross section and the
  $\mathrm{Z}\rightarrow 4\ell $ branching fraction, and constraints on
  anomalous triple gauge couplings at $\sqrt{s}=13\,\mathrm{TeV}$}'',} \textit{
  Eur. Phys. J. C} \textbf{ 78} (2018) 165,
  \href{http://dx.doi.org/10.1140/epjc/s10052-018-5567-9}{\doi{10.1140/epjc/s10052-018-5567-9}},
\href{http://www.arXiv.org/abs/1709.08601}{\texttt{arXiv:1709.08601}}.

\bibitem{Aad:2014wra}
\hrefCMSnoop {}{{ATLAS Collaboration}, ``{Measurements of four-lepton
  production at the \Z resonance in pp collisions at $\sqrt{s}=7$ and
  $8\,\mathrm{TeV}$ with ATLAS}'',} \textit{ Phys. Rev. Lett.} \textbf{ 112}
  (2014) 231806,
  \href{http://dx.doi.org/10.1103/PhysRevLett.112.231806}{\doi{10.1103/PhysRevLett.112.231806}},
\href{http://www.arXiv.org/abs/1403.5657}{\texttt{arXiv:1403.5657}}.

\bibitem{Aad:2015rka}
\hrefCMSnoop {}{{ATLAS Collaboration}, ``{Measurements of four-lepton
  production in pp collisions at $\sqrt{s}=$ 8 TeV with the ATLAS detector}'',}
  \textit{ Phys. Lett. B} \textbf{ 753} (2016) 552,
  \href{http://dx.doi.org/10.1016/j.physletb.2015.12.048}{\doi{10.1016/j.physletb.2015.12.048}},
\href{http://www.arXiv.org/abs/1509.07844}{\texttt{arXiv:1509.07844}}.

\bibitem{Acton:1991dq}
\hrefCMSnoop {}{{OPAL} Collaboration, ``{A measurement of photon radiation in
  lepton pair events from $\cPZ^0$ decays}'',} \textit{ Phys. Lett. B} \textbf{
  273} (1991) 338,
\href{http://dx.doi.org/10.1016/0370-2693(91)91694-Q}{\doi{10.1016/0370-2693(91)91694-Q}}.

\bibitem{Aad:2015sda}
\hrefCMSnoop {}{{ATLAS Collaboration}, ``{Search for Higgs and \Z boson decays
  to $\JPsi\gamma$ and $\Upsilon\mathrm{(nS)}\gamma$ with the ATLAS
  detector}'',} \textit{ Phys. Rev. Lett.} \textbf{ 114} (2015) 121801,
  \href{http://dx.doi.org/10.1103/PhysRevLett.114.121801}{\doi{10.1103/PhysRevLett.114.121801}},
\href{http://www.arXiv.org/abs/1501.03276}{\texttt{arXiv:1501.03276}}.

\bibitem{Aaboud:2017xnb}
\hrefCMSnoop {}{{ATLAS Collaboration}, ``{Search for exclusive Higgs and $\Z$
  boson decays to $\phi\gamma$ and $\rho\gamma$ with the ATLAS detector}'',}
  (2017).
  \href{http://www.arXiv.org/abs/1712.02758}{\texttt{arXiv:1712.02758}}.
Submitted to {\it JHEP}.

\bibitem{Bergstrom:1990uu}
\hrefCMSnoop {}{L.~Bergstr{\"o}m and R.~W. Robinett, ``{\Z decay into a vector
  meson and a lepton pair}'',} \textit{ Phys. Lett. B} \textbf{ 245} (1990)
  249,
\href{http://dx.doi.org/10.1016/0370-2693(90)90142-S}{\doi{10.1016/0370-2693(90)90142-S}}.

\bibitem{Fleming:1993fq}
\hrefCMSnoop {}{S.~Fleming, ``{Electromagnetic production of quarkonium in
  $\cPZ^0$ decay}'',} \textit{ Phys. Rev. D} \textbf{ 48} (1993) 1914,
  \href{http://dx.doi.org/10.1103/PhysRevD.48.R1914}{\doi{10.1103/PhysRevD.48.R1914}},
\href{http://www.arXiv.org/abs/hep-ph/9304270}{\texttt{arXiv:hep-ph/9304270}}.

\bibitem{Fleming:1994iu}
\hrefCMSnoop {}{S.~Fleming, ``{\PJGy production from electromagnetic
  fragmentation in $\cPZ^0$ decay}'',} \textit{ Phys. Rev. D} \textbf{ 50}
  (1994) 5808,
  \href{http://dx.doi.org/10.1103/PhysRevD.50.5808}{\doi{10.1103/PhysRevD.50.5808}},
\href{http://www.arXiv.org/abs/hep-ph/9403396}{\texttt{arXiv:hep-ph/9403396}}.

\bibitem{Li:2010xu}
\hrefCMSnoop {}{R.~Li and J.-X. Wang, ``{Next-to-leading-order QCD correction
  to inclusive $\PJGy(\Upsilon)$ production in $\cPZ^0$ decay}'',} \textit{
  Phys. Rev. D} \textbf{ 82} (2010) 054006,
  \href{http://dx.doi.org/10.1103/PhysRevD.82.054006}{\doi{10.1103/PhysRevD.82.054006}},
\href{http://www.arXiv.org/abs/1007.2368}{\texttt{arXiv:1007.2368}}.

\bibitem{Braaten:1993mp}
\hrefCMSnoop {}{E.~Braaten, K.~Cheung, and T.~C. Yuan, ``{$\cPZ^0$ decay into
  charmonium via charm quark fragmentation}'',} \textit{ Phys. Rev. D} \textbf{
  48} (1993) 4230,
  \href{http://dx.doi.org/10.1103/PhysRevD.48.4230}{\doi{10.1103/PhysRevD.48.4230}},
\href{http://www.arXiv.org/abs/hep-ph/9302307}{\texttt{arXiv:hep-ph/9302307}}.

\bibitem{Ernstrom:1996aa}
\hrefCMSnoop {}{P.~Ernstr{\"o}m, L.~L{\"o}nnblad, and M.~V{\"a}nttinen,
  ``{Evolution effects in $\cPZ^0$ fragmentation into charmonium}'',} \textit{
  Z. Phys. C} \textbf{ 76} (1997) 515,
  \href{http://dx.doi.org/10.1007/s002880050574}{\doi{10.1007/s002880050574}},
\href{http://www.arXiv.org/abs/hep-ph/9612408}{\texttt{arXiv:hep-ph/9612408}}.

\bibitem{Bergstrom:1990bu}
\hrefCMSnoop {}{L.~Bergstr{\"o}m and R.~W. Robinett, ``{On the rare decays $\Z
  \to \mathrm{V V}$ and $\Z \to \mathrm{V P}$}'',} \textit{ Phys. Rev. D}
  \textbf{ 41} (1990) 3513,
\href{http://dx.doi.org/10.1103/PhysRevD.41.3513}{\doi{10.1103/PhysRevD.41.3513}}.

\bibitem{PDG2016}
\hrefCMSnoop {}{{Particle Data Group}, ``Review of particle physics'',}
  \textit{ Chin. Phys. C} \textbf{ 40} (2016) 100001,
  \href{http://dx.doi.org/10.1088/1674-1137/40/10/100001}{\doi{10.1088/1674-1137/40/10/100001}}.

\bibitem{Doroshenko:1987nj}
\hrefCMSnoop {}{M.~N. Doroshenko, V.~G. Kartvelishvili, E.~G. Chikovani, and
  S.~M. Esakiya, ``{Vector quarkonium in decays of heavy Higgs particles}'',}
  \textit{ Yad. Fiz.} \textbf{ 46} (1987) 864.
[{\it Sov. J. Nucl. Phys.} \textbf{46} (1987) 493].

\bibitem{Gao:2014xlv}
\hrefCMSnoop {}{D.-N. Gao, ``{A note on Higgs decays into \Z boson and
  \PJGy($\Upsilon$)}'',} \textit{ Phys. Lett. B} \textbf{ 737} (2014) 366,
  \href{http://dx.doi.org/10.1016/j.physletb.2014.09.019}{\doi{10.1016/j.physletb.2014.09.019}},
\href{http://www.arXiv.org/abs/1406.7102}{\texttt{arXiv:1406.7102}}.

\bibitem{Gonzalez-Alonso:2014rla}
\hrefCMSnoop {}{M.~Gonzalez-Alonso and G.~Isidori, ``{The $h \to 4l$ spectrum
  at low $m_{34}$: Standard model vs. light new physics}'',} \textit{ Phys.
  Lett. B} \textbf{ 733} (2014) 359,
  \href{http://dx.doi.org/10.1016/j.physletb.2014.05.004}{\doi{10.1016/j.physletb.2014.05.004}},
\href{http://www.arXiv.org/abs/1403.2648}{\texttt{arXiv:1403.2648}}.

\bibitem{Chatrchyan:2012xi}
\hrefCMSnoop {}{{CMS Collaboration}, ``{Performance of CMS muon reconstruction
  in \Pp\Pp\ collision events at $\sqrt{s}=7\,\mathrm{TeV}$}'',} \textit{
  JINST} \textbf{ 7} (2012) P10002,
  \href{http://dx.doi.org/10.1088/1748-0221/7/10/P10002}{\doi{10.1088/1748-0221/7/10/P10002}},
\href{http://www.arXiv.org/abs/1206.4071}{\texttt{arXiv:1206.4071}}.

\bibitem{Khachatryan:2015hwa}
\hrefCMSnoop {}{{CMS Collaboration}, ``{Performance of electron reconstruction
  and selection with the CMS detector in proton-proton collisions at $\sqrt{s}
  = 8\,\mathrm{TeV}$}'',} \textit{ JINST} \textbf{ 10} (2015) P06005,
  \href{http://dx.doi.org/10.1088/1748-0221/10/06/P06005}{\doi{10.1088/1748-0221/10/06/P06005}},
\href{http://www.arXiv.org/abs/1502.02701}{\texttt{arXiv:1502.02701}}.

\bibitem{Chatrchyan:2008zzk}
\hrefCMSnoop {}{{CMS Collaboration}, ``The {CMS} experiment at the {CERN}
  {LHC}'',} \textit{ JINST} \textbf{ 3} (2008) S08004,
\href{http://dx.doi.org/10.1088/1748-0221/3/08/S08004}{\doi{10.1088/1748-0221/3/08/S08004}}.

\bibitem{Frixione:2007vw}
\hrefCMSnoop {}{S.~Frixione, P.~Nason, and C.~Oleari, ``{Matching NLO QCD
  computations with parton shower simulations: the \POWHEG method}'',} \textit{
  JHEP} \textbf{ 11} (2007) 070,
  \href{http://dx.doi.org/10.1088/1126-6708/2007/11/070}{\doi{10.1088/1126-6708/2007/11/070}},
\href{http://www.arXiv.org/abs/0709.2092}{\texttt{arXiv:0709.2092}}.

\bibitem{Sjostrand:2007gs}
\hrefCMSnoop {}{T.~Sj{\"o}strand, S.~Mrenna, and P.~Skands, ``{A brief
  introduction to PYTHIA 8.1}'',} \textit{ Comput. Phys. Commun.} \textbf{ 178}
  (2008) 852,
  \href{http://dx.doi.org/10.1016/j.cpc.2008.01.036}{\doi{10.1016/j.cpc.2008.01.036}},
\href{http://www.arXiv.org/abs/0710.3820}{\texttt{arXiv:0710.3820}}.

\bibitem{Sjostrand:2014zea}
T.~Sj{\"o}strand\hrefCMSnoop {}{ {et~al.}, ``An introduction to {PYTHIA
  8.2}'',} \textit{ Comput. Phys. Commun.} \textbf{ 191} (2015) 159,
  \href{http://dx.doi.org/10.1016/j.cpc.2015.01.024}{\doi{10.1016/j.cpc.2015.01.024}},
\href{http://www.arXiv.org/abs/1410.3012}{\texttt{arXiv:1410.3012}}.

\bibitem{Khachatryan:2015pea}
\hrefCMSnoop {}{{CMS Collaboration}, ``{Event generator tunes obtained from
  underlying event and multiparton scattering measurements}'',} \textit{ Eur.
  Phys. J. C} \textbf{ 76} (2016) 155,
  \href{http://dx.doi.org/10.1140/epjc/s10052-016-3988-x}{\doi{10.1140/epjc/s10052-016-3988-x}},
\href{http://www.arXiv.org/abs/1512.00815}{\texttt{arXiv:1512.00815}}.

\bibitem{Ball:2014uwa}
\hrefCMSnoop {}{{NNPDF} Collaboration, ``{Parton distributions for the LHC Run
  II}'',} \textit{ JHEP} \textbf{ 04} (2015) 040,
  \href{http://dx.doi.org/10.1007/JHEP04(2015)040}{\doi{10.1007/JHEP04(2015)040}},
\href{http://www.arXiv.org/abs/1410.8849}{\texttt{arXiv:1410.8849}}.

\bibitem{Agostinelli:2002hh}
\hrefCMSnoop {}{{G{\sc eant4}} Collaboration, ``G{\sc eant4} --- a simulation
  toolkit'',} \textit{ Nucl. Instrum. Meth. A} \textbf{ 506} (2003) 250,
\href{http://dx.doi.org/10.1016/S0168-9002(03)01368-8}{\doi{10.1016/S0168-9002(03)01368-8}}.

\bibitem{Cacciari:2008gp}
\hrefCMSnoop {}{M.~Cacciari, G.~P. Salam, and G.~Soyez, ``The anti-\kt jet
  clustering algorithm'',} \textit{ JHEP} \textbf{ 04} (2008) 063,
  \href{http://dx.doi.org/10.1088/1126-6708/2008/04/063}{\doi{10.1088/1126-6708/2008/04/063}},
  \href{http://www.arXiv.org/abs/0802.1189}{\texttt{arXiv:0802.1189}}.

\bibitem{Cacciari:2011ma}
\hrefCMSnoop {}{M.~Cacciari, G.~P. Salam, and G.~Soyez, ``{FastJet} user
  manual'',} \textit{ Eur. Phys. J. C} \textbf{ 72} (2012) 1896,
  \href{http://dx.doi.org/10.1140/epjc/s10052-012-1896-2}{\doi{10.1140/epjc/s10052-012-1896-2}},
\href{http://www.arXiv.org/abs/1111.6097}{\texttt{arXiv:1111.6097}}.

\bibitem{Barlow:1990vc}
\hrefCMSnoop {}{R.~J. Barlow, ``{Extended maximum likelihood}'',} \textit{
  Nucl. Instrum. Meth. A} \textbf{ 297} (1990) 496,
\href{http://dx.doi.org/10.1016/0168-9002(90)91334-8}{\doi{10.1016/0168-9002(90)91334-8}}.

\bibitem{ref:FisherComb}
\hrefCMSnoop {}{R.~A. Fisher, ``{Questions and Answers $\#14$}'',} \textit{
  Amer. Statis.} \textbf{ 2} (1948) 30,
  \href{http://dx.doi.org/10.2307/2681650}{\doi{10.2307/2681650}}.

\bibitem{Oreglia:1980cs}
\href {http://www.slac.stanford.edu/cgi-wrap/getdoc/slac-r-236.pdf}{M.~J.
  Oreglia, ``A study of the reactions $\psi^\prime \to \gamma \gamma \psi$''}.
\newblock PhD thesis, Stanford University, 1980.
\newblock {SLAC} Report {SLAC-R-236}.

\bibitem{Khachatryan:2010xn}
\hrefCMSnoop {}{{CMS Collaboration}, ``{Measurements of inclusive $\mathrm{W}$
  and $\Z$ cross sections in $\mathrm{pp}$ collisions at
  $\sqrt{s}=7\,\mathrm{TeV}$}'',} \textit{ JHEP} \textbf{ 01} (2011) 080,
  \href{http://dx.doi.org/10.1007/JHEP01(2011)080}{\doi{10.1007/JHEP01(2011)080}},
\href{http://www.arXiv.org/abs/1012.2466}{\texttt{arXiv:1012.2466}}.

\bibitem{Pivk:2004ty}
\hrefCMSnoop {}{M.~Pivk and F.~R. Le~Diberder, ``{$_{s}\mathcal{P}lot$: A
  statistical tool to unfold data distributions}'',} \textit{ Nucl. Instrum.
  Meth. A} \textbf{ 555} (2005) 356,
  \href{http://dx.doi.org/10.1016/j.nima.2005.08.106}{\doi{10.1016/j.nima.2005.08.106}},
\href{http://www.arXiv.org/abs/physics/0402083}{\texttt{arXiv:physics/0402083}}.

\bibitem{Faccioli:2010kd}
\hrefCMSnoop {}{P.~Faccioli, C.~Louren\c{c}o, J.~Seixas, and H.~K. W{\"o}hri,
  ``{Towards the experimental clarification of quarkonium polarization}'',}
  \textit{ Eur. Phys. J. C} \textbf{ 69} (2010) 657,
  \href{http://dx.doi.org/10.1140/epjc/s10052-010-1420-5}{\doi{10.1140/epjc/s10052-010-1420-5}},
\href{http://www.arXiv.org/abs/1006.2738}{\texttt{arXiv:1006.2738}}.

\end{thebibliography}\endgroup
